\documentclass{article}

\usepackage{amsmath}
\usepackage{amsfonts}
\usepackage{amssymb}

\def\be{\begin{eqnarray}}
\def\ee{\end{eqnarray}}
\def\nn{\nonumber}

\def\co{:}

\hoffset= - 1.0in         % switch off for draft style
\title{{\bf The Power of Nekrasov Functions
} \vspace{.2cm}}
\author{{\bf A.Mironov}\footnote{ {\small {\it
Lebedev Physics Institute} and {\it ITEP, Moscow, Russia}};
mironov@itep.ru; mironov@lpi.ru} \ and {\bf
A.Morozov}\thanks{{\small {\it ITEP, Moscow, Russia}};
morozov@itep.ru} \date{ }}

\begin{document}

\maketitle

\vspace{-5.0cm}

\begin{center}
\hfill FIAN/TD-18/09\\
\hfill ITEP/TH-34/09\\
\end{center}

\vspace{3.cm}

\begin{abstract}
\noindent The recent AGT suggestion \cite{AGT} to use the set of
Nekrasov functions \cite{Nek} as a basis for a linear decomposition
of generic conformal blocks works very well not only in the case of
Virasoro symmetry, but also for conformal theories with extended
chiral algebra. This is rather natural, because Nekrasov functions
are introduced as expansion basis for generalized hypergeometric
integrals, very similar to those which arise in expansion of
Dotsenko-Fateev integrals in powers of alpha-parameters. Thus, the
AGT conjecture is closely related to the old belief that conformal
theory can be effectively described in the free field formalism, and
it can actually be a key to clear formulating and proof this
long-standing hypothesis. As an application of this kind of
reasoning we use knowledge of the exact hypergeometric conformal block
for complete proof of the AGT relation for a restricted
class of external states.
\end{abstract}

\section{Introduction}

Along with the matrix-model $\tau$-functions \cite{Mtau}, the
Nekrasov functions \cite{Nek}, being coefficients of
character expansions of the former ones \cite{charexp}, are very important
new special functions, badly needed
for developing a quantitative string theory \cite{AMMsf}. They
originally appeared in the framework of the instanton expansion of Seiberg-Witten
quasi-classical (Whitham) $\tau$-functions \cite{SW} and provide
a kind of a quantization of Seiberg-Witten prepotential
\cite{NeSha}. These functions with the theory of
Hurwitz-Kontsevich functions \cite{HuKo}, describing combinatorics
of ramified Riemann surfaces, an essential subject for
perturbative and non-perturbative string theory \cite{rams}. A new
important application was recently suggested in \cite{AGT}: generic
Virasoro conformal blocks \cite{BPZ}-\cite{DiF} can be nicely
represented as linear combinations of Nekrasov functions for the $U(2)$
quiver models \cite{quiv}. For technical details of this AGT
relation see \cite{Wyl,MMMagt}. The AGT suggestion has a number of
natural generalizations, the first in the line being that to $SU(N)$
models, which, on the conformal side, should correspond to theories with
the conformal algebra extended from Virasoro to $W^{(N)}$. The most
straightforward idea in this direction would be to simply decompose
certain $W^{(N)}$ conformal blocks as combinations of $U(N)$-quiver
Nekrasov functions \cite{Wyl}. This idea works perfectly well for
the "perturbative" (quasiclassical) Nekrasov functions which coincide not only
with the DOZZ triple vertices \cite{DOZZ} in Liouville model ($U(2)$-case), but
also with the Fateev-Litvinov vertices \cite{FLit} in affine Toda
models ($U(N)$-case). It is now checked up to level two (order $x^2$) in the
$W^{(3)}$ conformal blocks \cite{mmAGT}, where calculation depends
on some knowledge about $W^{(3)}$ conformal blocks (we refer for details
to a dedicated elementary-level summary in \cite{MMMM}).

In this short note we want to attract attention to another aspect
of the AGT proposal. The Nekrasov integrals can be considered as an appropriate
analytical continuation of expansion coefficients of
the Dotsenko-Fateev integrals ("screening charges") \cite{FD} in
powers of $\alpha$-parameters.
Thus, an apparent success
of the AGT conjecture in description of generic conformal blocks
can be considered as a strong support of the old hypothesis
that the free field formalism can indeed be used
to describe generic conformal theories.
This is well established in particular distinguished examples
\cite{FD,GMMOS,FZ}, but in the check of the AGT relations
in \cite{AGT,Wyl,MMMagt,mmAGT} one actually works without
any reference to particular model, only to its conformal
properties.
In what follows, we briefly list a small set of examples,
which show how expansion in Nekrasov functions generalizes
the standard hypergeometric series to the ones needed
in description of the Dotsenko-Fateev integrals.
Transition to generic conformal blocks still looks mysterious.
However, now it can be formulated in a very clear and
general form, and further work on the AGT relations would presumably
clarify the old mysteries of representation theory of
chiral algebras and make the subject much more transparent and
understandable.

\newpage

\unitlength 1mm % = 2.845pt
\linethickness{0.4pt}
\ifx\plotpoint\undefined\newsavebox{\plotpoint}\fi % GNUPLOT compatibility
\begin{picture}(47.94,-06.98)(-110,42)
\put(18.061,18.879){\line(1,0){22.818}}
%\emline(8.993,24.899)(17.838,18.953)
\multiput(8.993,24.899)(.049970214,-.033593421){177}{\line(1,0){.049970214}}
%\end
%\emline(17.838,18.953)(9.142,13.973)
\multiput(17.838,18.953)(-.058757277,-.03364733){148}{\line(-1,0){.058757277}}
%\end
%\emline(47.94,24.899)(40.953,18.879)
\multiput(47.94,24.899)(-.039031239,-.033633302){179}{\line(-1,0){.039031239}}
%\end
%\emline(40.953,18.879)(47.94,13.899)
\multiput(40.953,18.879)(.047206701,-.03364733){148}{\line(1,0){.047206701}}
%\end
\put(9.96,27){$\vec\alpha_1$}
\put(9.96,10){$\vec\alpha_2$}
\put(47.048,27){$\vec\alpha_3$}
\put(47.048,10){$\vec\alpha_4$}
\put(28.392,20.662){$\vec\alpha$}
\end{picture}
\section{Conformal blocks and their expansions}

\bigskip

Conformal blocks often have a pronounced form of hypergeometric series.

\bigskip

\noindent
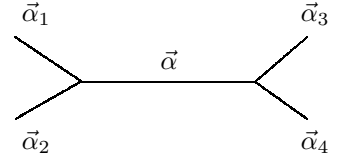
\underline{Free field conformal block:}
\be
B(x) = (1-x)^{-2\vec\alpha_1\vec\alpha_3} = (1-x)^{-A}
= 1 + A\,x + \frac{A(A+1)}{2}\,x^2 + \frac{A(A+1)(A+2)}{6}\,x^3 + \ldots
\ee

\noindent
\underline{Fateev-Litvinov conformal block for one special and one
maximally degenerate states at external lines:}
In \cite{FLit} it was shown that generic hypergeometric series is represented by
subset of conformal blocks of the $SL(N)$ Toda model with restricted external states:
\be\label{FL}
B(x) = \phantom._{N}F_{N-1}(A_1,...,A_N;B_1,...,B_{N-1};x)
= 1 + x\,\frac{A_1...A_N}{B_1...B_{N-1}} + \frac{x^2}{2}\,
\frac{A_1(A_1+1)...A_N(A_N+1)}{B_1(B_1+1)...B_{N-1}(B_{N-1}+1)}
+ \ldots
\ee
The free field case is reproduced when $A_i=B_i$, $A_N=A$.

\bigskip

Thus, an arbitrary hypergeometric function is some conformal block.
However, inverse is not true:
the generic conformal block does not have this simple
hypergeometric structure.

\bigskip

\noindent
\underline{Dotsenko-Fateev integral:}\footnote{
We denote the screening charge parameter $\epsilon_-$
rather than $\epsilon_+$, because
after rescaling of dimensions \cite{MMMagt}
$\Delta\rightarrow \Delta/(\epsilon_+\epsilon_-)$ the
Gamma-functions acquire the form
$\Gamma\left(\frac{2\alpha_3\epsilon_-}{\epsilon_+\epsilon_-}+n\right)\sim
2\alpha_3(2\alpha_3+\epsilon_+)\ldots \big(2\alpha_3+(n-1)\epsilon_+\big)$,
which should be compared with the {\it chiral}
Nekrasov function (\ref{Zchi}). The screenings with $\epsilon_+$
correspond to {\it anti-chiral} functions.
}
\be
B(x) =
\left< \co e^{\alpha_1\phi(1)}\co\ \co e^{\alpha_2\phi(0)}\co\
\co e^{\alpha_3\phi(x)}\co\ \co e^{\alpha_4\phi(\infty)}\co
\oint  \co e^{\epsilon_-\phi(z)}dz\co \right> \ \sim\nn \\
\sim\ (1-x)^{-2\alpha_1\alpha_3}\int_0^1 z^{-2\alpha_2\epsilon_-}
(1-z)^{-2\alpha_3\epsilon_-}(z-x)^{2\epsilon_-\epsilon_-}dz
\ \sim\ (1-x)^{-A}\int_0^1 z^{-B}(1-z)^{-C}(z-x)^{-D}dz\ \sim\nn\\
\sim\ (1-x)^{-A}\sum_n \frac{x^n}{n!}\cdot
\frac{\Gamma(D+n)\Gamma(B+C+D-1+n)}{\Gamma(B+D+n)}
= (1-x)^{-A}\cdot\phantom._2F_1(D,B+C+D-1;B+D;x)
\label{FDint}
\ee

\bigskip

\noindent
\underline{Multiple Dotsenko-Fateev integrals:}\ \ \
when several screenings are inserted, one obtains instead of (\ref{FDint})
\be
B(x) = (1-x)^{-A}\oint\ldots\oint \prod_i z_i^{-B_i}(1-z_i)^{-C_i}
(z_i-x)^{-D_i}\prod_{i<j}(z_i-z_j)^{E_{ij}}\prod_i dz_i
\label{gehy}
\ee
In variance with (\ref{FDint}) such integrals are no longer
hypergeometric, but they are similar in some respects and are
often called
"generalized hypergeometric integrals", see for example \cite{SheVa}.
Still, if we are interested in expansion bases, the
terminological does not help: such $B(x)$ is {\it not} an
ordinary hypergeometric series of the type $\phantom._pF_q$,
unless the matrix $E_{ij}$ is very special
(roughly, $E_{ij}\sim E_i\delta_{i+1,j}$, see the last paper in
\cite{SheVa}). It also deserves mentioning that
there is a certain difference between these integrals for
a single free field and for multiple ($r=N-1$) fields: the less fields,
the more relations between the numerous parameters $A,B_i,C_i,D_i,E_{ij}$:
they are all made from $\epsilon_\pm$ and four $(N-1)$-component vectors
$\vec\alpha_1,\ldots,\vec\alpha_4$.\footnote{
In Dotsenko-Fateev approach, the dependence on the fifth vector $\vec\alpha$
is presumably restored from a sophisticated analytical continuation
of the answer with arbitrary number of screening insertions.
An exact relation between the multiple Fateev-Dotsenko integrals and
the Virasoro or $W$-conformal blocks
is believed to exist, but remains uncovered.
The AGT relation could be a key to resolve it.
}
Increasing $N$, one actually
enlarges class of the Fateev-Dotsenko integrals, they span the entire
space of generalized hypergeometric series only for $N\rightarrow\infty$.

\bigskip

\noindent
\underline{Virasoro case without external fields \cite{MMMagt}:}
\be
1 + x\cdot \frac{\Delta}{2}
+ \frac{x^2}{2}\cdot\frac{\Delta
\Big(8\Delta^3+(c+8)\Delta^2+(2c-8)\Delta+c\Big)}
{2\Big(16\Delta^2+(2c-10)\Delta + c\Big)}
+ \frac{x^3}{6}\cdot\frac{\Delta(\Delta+2)
\Big(8\Delta^3+(c+18)\Delta^2+(3c-14)\Delta+2c\Big)}
{4\Big(16\Delta^2+(2c-10)\Delta +c\Big)} + \ldots
\nn
\ee
In particular, for $c=1$:
\be
B(x) = 1+x\cdot \frac{\Delta}{2} + \frac{x^2}{2}\cdot
\frac{\Delta(8\Delta^3+9\Delta^2-6\Delta+1)}{2(4\Delta-1)^2} +
\frac{x^3}{6}\cdot\frac{\Delta(\Delta+2)(8\Delta^3+19\Delta^2-11\Delta+2)}
{4(4\Delta-1)^2} + \ldots
\label{CBvirc1}
\ee

\bigskip

\noindent
\underline{$W^{(3)}$ case, general central charge $c$, two states
$\vec\alpha_1$ and $\vec\alpha_3$ are {\it special} \cite{mmAGT}:}
\be
1 + x \, \frac{\frac{D\Delta}{2}{\cal D}_{12}{\cal D}_{34}
- \frac{w}{3}
\Big({\cal D}_{12}{\cal W}_{34} + {\cal W}_{12}{\cal D}_{34}\Big)
+ \frac{2\Delta}{9}{\cal W}_{12}{\cal W}_{34}}
{D\Delta^2-w^2} + O(x^2)
\ee
\be
{\cal D}_{12} = \Delta+\Delta_1-\Delta_2,
& {\cal D}_{34}=\Delta+\Delta_3-\Delta_{34} \nn\\
{\cal W}_{12} = w+2w_1-w_2+\frac{3w_1}{\Delta_1}(\Delta-\Delta_1-\Delta_2),
&\ \ \
{\cal W}_{34} = w+w_3+w_4-\frac{3w_3}{2\Delta_3}(\Delta+\Delta_3-\Delta_4)
\ee
Note numerous differences between ${\cal W}_{34}$ and ${\cal W}_{12}$
in signs and coefficients, see \cite{MMMM}.
The parameter $D = \Delta +\frac{3\epsilon^2}{4}$
and the central charge $c= 2(1-12\epsilon^2)$.

\bigskip

\noindent
\underline{$W^{(3)}$ case, no external states, $c=2$ \cite{mmAGT}:}
\be
B(x)=
1 + x\cdot\frac{\Delta\left(\Delta^3-\frac{8}{9}w^2\right)}{2(\Delta^3-w^2)}
+ \frac{x^2}{2}\cdot \frac{
81\Delta^4(\Delta-1)^4\big(8\Delta^3+9\Delta^2-6\Delta+1\big)
+ G_2w^2 + G_4w^4 + G_6w^6 }
{2\cdot 81\cdot (\Delta^3-w^2)\Big((4\Delta-1)(\Delta-1)^2-4w^2\Big)^2}
+\ldots
\label{CBw3c2}
\ee
\be
G_2= -72\Delta(\Delta-1)^2(25\Delta^5+20\Delta^4-25\Delta^3
+23\Delta^2-8\Delta+1),\nn
\ee
\be
G_4=1664\Delta^5+608\Delta^4-2688\Delta^3+2704\Delta^2
-608\Delta+48,\nn
\ee
\be
G_6 = -128\Delta(4\Delta+7)\nn
\ee
For $w=0$ this expression reduces to (\ref{CBvirc1}),
note that this happens despite $c=2$ in (\ref{CBw3c2}), while $c=1$ in
(\ref{CBvirc1}).
For $c\neq 2$ the level one term becomes
$\displaystyle{x\frac{\Delta\left(D\Delta^2-\frac{8}{9}w^2\right)}{2(D \Delta^2-w^2)}}$,
a similar deformation occurs in the second term, see \cite{mmAGT}.

Formulas (\ref{CBvirc1}) and (\ref{CBw3c2})
do not have the free field limit,
because for these values of central charge in the free field model
the intermediate states have vanishing
$\Delta$ and $w$ whenever all the external momenta are zero.

\bigskip

To summarize, ordinary hypergeometric functions $pF_q$
are not sufficient to describe arbitrary conformal blocks
(at least, the poorly studied
hypergeometric integrals \cite{SheVa} are needed),
and it is a natural question what should be a reasonable
extension of this class of functions to serve these purposes.
The AGT conjecture is actually a claim that the Nekrasov functions
can be an answer to this challenge.

\section{Nekrasov functions for ordinary Young diagrams}

The Nekrasov functions are defined as coefficients of the
character-like expansion of Nekrasov integrals,
very similar to those in (\ref{gehy}).
They are defined for the $N$-plets of Young diagrams.
For our purposes we consider only 1-point quiver functions
associated with $N_f=2N$ fundamentals.
Extension to other representations and quivers can be
immediately provided.
In the next section, we will also put
$\epsilon\equiv\epsilon_++\epsilon_-=0$
and $\epsilon_+=-\epsilon_-=1$. According to the AGT rules
this  corresponds to putting $c=1$ and $c=2$ in the Virasoro
and $W^{(3)}$ cases respectively. In this section we, however,
keep $\epsilon$ arbitrary.

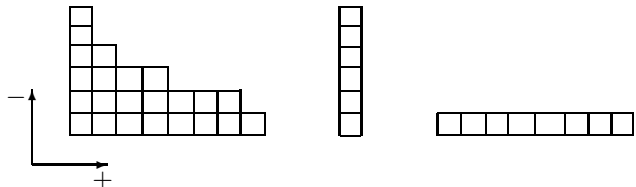
\begin{figure}[b]
\unitlength 1mm % = 2.845pt
\linethickness{0.4pt}
\ifx\plotpoint\undefined\newsavebox{\plotpoint}\fi % GNUPLOT compatibility
\begin{picture}(58.82,15.963)(-30,5)
%\emline(14.085,8.935)(39.943,8.829)
\put(14.085,8.935){\line(1,0){25.8576}}
%\end
%\emline(62.962,8.935)(88.82,8.829)
\put(62.962,8.935){\line(1,0){25.8576}}
%\end
\put(39.943,8.829){\line(0,1){3.048}}
\put(88.82,8.829){\line(0,1){3.048}}
\put(39.943,11.878){\line(-1,0){25.963}}
\put(88.82,11.878){\line(-1,0){25.963}}
\put(13.98,11.878){\line(0,1){14.085}}
\put(49.823,11.878){\line(0,1){14.085}}
\put(13.98,25.963){\line(1,0){2.943}}
\put(49.823,25.963){\line(1,0){2.943}}
\put(16.923,25.963){\line(0,-1){17.028}}
\put(52.766,25.963){\line(0,-1){17.028}}
%\emline(14.085,14.821)(36.684,14.926)
\put(14.085,14.821){\line(1,0){22.5991}}
%\end
\put(36.684,14.926){\line(0,-1){5.886}}
\put(33.636,14.821){\line(0,-1){5.781}}
\put(30.482,14.716){\line(0,-1){5.571}}
\put(14.19,17.974){\line(1,0){12.824}}
\put(27.014,17.974){\line(0,-1){8.935}}
\put(14.19,20.917){\line(1,0){5.991}}
\put(20.182,20.917){\line(0,-1){11.878}}
\put(23.65,17.869){\line(0,-1){8.829}}
%\emline(14.085,23.44)(16.818,23.335)
\put(14.085,23.44){\line(1,0){2.7329}}
%\end
%\emline(49.928,23.44)(52.661,23.335)
\put(49.928,23.44){\line(1,0){2.7329}}
%\end
\put(13.98,11.773){\line(0,-1){2.838}}
\put(49.823,11.773){\line(0,-1){2.838}}
\put(62.857,11.773){\line(0,-1){2.838}}
\put(49.928,8.829){\line(1,0){2.628}}
\put(49.928,11.878){\line(1,0){2.733}}
\put(49.823,14.821){\line(1,0){2.838}}
\put(49.928,17.974){\line(1,0){2.628}}
\put(49.928,20.602){\line(1,0){2.628}}
%\put(19.551,4.31){Diagram $Y = [112478]$}
%\put(50.874,4.204){$Y=[6]$}
%\put(75.681,4.625){chain diagram $Y=[11111111] = [1^8]$}
%\put(50.979,6.517){column diagram}
\put(85.982,11.667){\line(0,-1){2.733}}
\put(82.933,11.878){\line(0,-1){2.943}}
\put(79.78,11.773){\line(0,-1){2.838}}
\put(75.891,11.667){\line(0,-1){2.733}}
\put(66.01,11.773){\line(0,-1){2.838}}
\put(69.269,11.773){\line(0,-1){2.838}}
\put(72.212,11.773){\line(0,-1){2.838}}
\put(9,5){\vector(1,0){10}}
\put(9,5){\vector(0,1){10}}
\put(17,2){$+$}
\put(5.5,13){$-$}
\end{picture}
\caption{\footnotesize{ Three examples of Young diagrams:
a generic type diagram $Y = [64332221]$, a column $Y=[6]$
and a chain $Y=[11111111] = [1^8]$.
With the chains and columns are associated
the {\it chiral} and {\it anti-chiral} Nekrasov functions
respectively.
Horizontal and vertical directions are called $+$ and $-$,
because the "natural" position of the Young diagram is rotated by
$45^\circ$ counterclockwise.
}}
\end{figure}

The general definition of the Nekrasov function in the $SU(N)$ case is as follows
\be
Z(Y_1,\ldots,Y_N) = \frac{\prod_i \eta(a_i,Y_i)}
{\prod_{i,j} \xi(a_i-a_j,Y_i,Y_j)}
\ee
where
\be
\eta(a,Y) = \prod_{(p,q)\in Y} P\Big(a+(p-1)\epsilon_+ + (q-1)\epsilon_-\Big)
\label{etadef}
\ee
for a polynomial $P(a) = \prod_{f=1}^{N_f=2N} (a_i+\mu_f)$,
and $\xi$ is a comprehensive deformation of the hook
formula for $d_Y^{-2}=\xi(0,Y,Y)$ with $\epsilon_-=-\epsilon_+=-1$:
\be
\xi(a,Y_1,Y_2)
= -\prod_{(p,q)\in Y_1}
\Big(a+\epsilon_+\left(k^T_j(Y_1)-i+1\right) - \epsilon_-
\left(k_i(Y_2) - j\right)\Big)
\Big(a+\epsilon_+\left(k^T_j(Y_1)-i\right) - \epsilon_-
\left(k_i(Y_2) - j+1\right)\Big)
\ee
Here $k_i(Y)$ is the height of the $i$-th column and
$k_j^T(Y)$ is the length of the $j$-th row of the diagram
$Y$, which is also denoted as $Y=[k_1k_2\ldots]$.
Of course $k_1\geq k_2\geq\ldots\geq 0$ and
$k_1^T\geq k_2^T\geq \ldots 0$.
Note that the product runs over the first diagram only, while the second diagram
enters just through the coefficient $k_i(Y_2)$ in the product. Therefore, it is not
surprising that when the second diagram is empty, this expression simplifies to
$\xi(a,\emptyset,Y)=1$, while the first diagram is empty, the expression reduces to
\be
\xi(a,Y,\emptyset) = \prod_{(p,q)\in Y}
\Big(a + (p-1)\epsilon_+ + (q-1)\epsilon_-\Big)\Big(a + p\epsilon_+ + q\epsilon_-\Big)
\ee
so that for fixed $i$
\be
\prod_{j\neq i}\xi(a_i-a_j,Y_i,\emptyset) =
\prod_{(p,q)\in Y_i} Q_i\Big(a_i+(p-1)\epsilon_+ + (q-1)\epsilon_-\Big)
\ee
which is very  similar to (\ref{etadef}), only with $P(a)$
of degree $N_f=2N$ substituted by the polynomial
\be
Q_i(x) = \prod_{j\neq i} (x-a_j)(x-a_j+\epsilon)
\ee
of degree $2N-2$.
Further  simplification occurs when one restricts $Y$ to be either
a line or a column.

These {\it chiral} Nekrasov functions have a clear form
of the hypergeometric series terms:
\be
Z(\emptyset\ldots [1^n]\ldots\emptyset)=
\frac{1}{\epsilon_-\epsilon_+^nn!}
\frac{P(a_i)P(a_i+\epsilon_+)\ldots P\big(a_i+(n-1)\epsilon_+\big)}
{(-\epsilon_-)(-\epsilon_-+\epsilon_+)...(-\epsilon_-+(n-1)\epsilon_+)
Q_i(a_i)Q_i\big(a_i+\epsilon_+\big)\ldots
Q_i\big(a_i+(n-1)\epsilon_+\big)}
\label{Zchi}
\ee
where the diagram $Y_i=[1^n]$ stands on the $i$-th place in the $N$-plet and
the additional $\epsilon$-dependent factor comes from $\xi(0,Y_i,Y_i)$.

The {\it anti-chiral} functions $Z(\emptyset\ldots [n]\ldots\emptyset)$
have exactly the same form with $\epsilon_+\rightarrow \epsilon_-$.

\section{AGT relations}

\subsection{Fateev-Litvinov conformal block via
Nekrasov functions}

The polynomial $P(a)$ can be adjusted so that only one diagram
contributes at each level:
\be
P(a_j) = 0 \ \ \ \ {\rm for} \ \ j\neq i\ \ \ \
{\rm and}\ \ \ \  P(a_i+\epsilon_-)=0
\ee
this fixes $N$ out of $N_f=2N$ parameters $\mu_f$.
The first condition is needed to eliminate all
$N$-plets of Young diagrams, where any non-empty diagram
stands at any position, different from $i$, including all
mixed diagrams, where at least two non-trivial diagrams are present
in the $N$-plet.
The second condition eliminates in addition all diagrams
except for the single-row $[1^n]$, standing in
the $i$-th position in the $N$-plet.
Both kind of requirements should be clear from a look at (\ref{etadef}).

In this way, one can describe various hypergeometric series
by the Fateev-Litvinov conformal blocks (\ref{FL})
which depend on exactly $2N$
free parameters: $\vec\alpha_2,\vec\alpha_4,\alpha_1$ and $c$,
all combined in a sophisticated way into
$A_1,...,A_N,B_1,...,B_{N-1}$.
Similarly to the free field case, the intermediate-state momentum
$\vec\alpha$ for these specific external states
is almost (up to N possible values) dictated by the external momenta
due to severe selection rules of the Toda-chain model).
{\bf This argument provides a complete proof of the AGT relation
in this restricted setting} (such a possibility has been also
anticipated in \cite{Wyl}).

\subsection{Deviations from chirality and hypergeometricity}

The non-chiral $SU(2)$ Nekrasov functions can be used to introduce
the needed corrections for the Virasoro conformal block,
which deviates it from the hypergeometric form.
For the Dotsenko-Fateev integrals it is almost obvious,
if one recalls that the Nekrasov integrals \cite{Nek} ($P(x)$ was defined in s.3)
\be
Z_k^{SU(N)}=\int\prod_{i=1}^k{dz_i\over 2\pi i}{1\over k!}\left(
{\epsilon\over\epsilon_+\epsilon_-}\right)^k
{\Delta (0)\Delta (\epsilon)\over\Delta (\epsilon_+)\Delta(\epsilon_-)}
{P(z_i)\over Q(z_i)},
\\ \ \ \ \ \Delta(x)\equiv \prod_{i<j}\Big((z_i-z_j)^2-x^2\Big),
\ \ \ \ \ \ Q(x)\equiv\prod_i (x-a_i)(x-a_i+\epsilon),
\ee
are nothing but the generalized hypergeometric integrals \cite{SheVa}
which appeared in (\ref{gehy}).
It is not that clear what they have to do with the series
like (\ref{CBvirc1}) and (\ref{CBw3c2}) -- but they do,
as successful tests of the AGT relation clearly demonstrate,
and this is in perfect agreement with beliefs in
conformal field theory and representation theory of loop
algebras, where conformal blocks are thought to be somehow
expandable in analytically continued Dotsenko-Fateev integrals.
As it is now obvious, the further character-like
expansion of integrals in the
Nekrasov functions converts the problem into the
very clearly formulated AGT conjecture,
and hopefully provides a key for its final resolution.

\subsection{$SU(2)$/Virasoro case}

The $SU(2)$ case illustrates nicely that
including the non-chiral Nekrasov function allows one to
extend hypergeometric series into an interesting direction:
for example, to describe the Virasoro conformal blocks.

Let us see how it works at the first two levels \cite{MMMagt}:
$$
Z^{SU(2)}(x) = 1 + x\Big(Z(\Box,\emptyset)+Z(\emptyset,\Box)\Big)
+ x^2\Big(\underline{Z(\Box\Box,\emptyset) + Z(\emptyset,\Box\Box)}
+\underline{\underline{
Z({\Box\over\Box},\emptyset) + Z(\emptyset,{\Box\over \Box})}}
+ Z(\Box,\Box)\Big) + \ldots =
$$
\vspace{-0.5cm}
\be
= 1 + x\cdot Z_1^{SU(2)} + x^2\cdot Z_2^{SU(2)} + \ldots
= 1 + x\! \left(\frac{P(a)}{Q(2a)}+\frac{P(-a)}{Q(-2a)}\right)+
\label{NekSU2}
\ee
\vspace{-0.3cm}
$$
+ x^2\!\left(\underline{\frac{P(a)P(a+1)}{4Q(2a)Q(2a+1)}
+ \frac{P(-a)P(-a+1)}{4Q(-2a)Q(-2a+1)} } +
\underline{\underline{\frac{P(a)P(a-1)}{4Q(2a)Q(2a-1)}
+ \frac{P(-a)P(-a-1)}{4Q(-2a)Q(-2a-1)} }}
+ \frac{P(a)P(a)}{Q(2a-1)Q(2a+1)}
\right)
$$
Here $a_2=-a_1=-a$ and we denote $Q_1(a_1) = Q(2a)$, $Q_2(a_2)=Q(-2a)$.
Underlined and double-underlined are the chiral
and anti-chiral
contributions, non-underlined remains the mixing contribution
from the pair of diagrams.
At level one, there is no difference between the chiral and anti-chiral
diagrams.

Matching with (\ref{CBvirc1}) is achieved if one takes
$P(a)=a^4$, $Q(2a)$ = $(2a)^2$:
\be
Z = 1 + x\cdot\frac{a^2}{2} + {x^2}\cdot
\left(2\frac{a^4(a+1)^4}{16a^2(2a+1)^2} + 2\frac{a^4(a-1)^4}{16a^2(2a-1)^2}
+ \frac{a^8}{(2a-1)^2(2a+1)^2}\right) + \ldots =\nn\\
= 1 + x\cdot\frac{a^2}{2} + \frac{x^2}{2}\cdot
\frac{a^2\Big((a+1)^4(2a-1)^2+(a-1)^2(2a+1)^2 +8a^6\Big)}{4(4a^2-1)^2}
+ \ldots = \nn \\
= 1 + x\cdot\frac{a^2}{2} + \frac{x^2}{2}\cdot
\frac{a^2(8a^6+9a^4-6a^2+1)}{2(4a^2-1)^2} + \ldots
\ee
If one now identifies $\Delta=a^2$, then this expression
reproduces (\ref{CBvirc1}).
It is straightforward to generalize this calculation to
arbitrary central charges and external states, to level $3$
\cite{MMMagt} and further \cite{AGT}.

\subsection{$SU(3)/W^{(3)}$ case
\label{SU3prob}}

Instead of (\ref{NekSU2}) one now has
$$
Z^{SU(3)} = 1 + x\Big(Z(\Box,\emptyset,\emptyset)
+Z(\emptyset,\Box,\emptyset)
+Z(\emptyset,\emptyset,\Box)\Big)
+ x^2\Big(\underline{Z(\Box\Box,\emptyset,\emptyset)
+ Z(\emptyset,\Box\Box,\emptyset)
+ Z(\emptyset,\emptyset,\Box\Box) } +
$$ $$
+\underline{\underline{
Z({\Box\over\Box},\emptyset,\emptyset)
+ Z(\emptyset,{\Box\over\Box},\emptyset)
+ Z(\emptyset,\emptyset,{\Box\over\Box})  }}
+ Z(\Box,\Box,\emptyset) + Z(\Box,\emptyset,\Box) + Z(\emptyset,\Box,\Box)
 \Big) + \ldots = 1 + xZ_1 + x^2 Z_2 + \ldots =
$$
\vspace{-0.5cm}
\be
= 1 + x \left(\frac{P(a_1)}{Q(a_1)}+\frac{P(a_2)}{Q(a_2)}
+ \frac{P(a_3)}{Q(a_3)}\right)+
\label{NekSU3}
\ee
\vspace{-0.3cm}
$$
+ x^2\left(\underline{\frac{P(a_1)P(a_1+1)}{4Q(a_1)Q(a_1+1)}
+ \frac{P(a_2)P(a_2+1)}{4Q(a_2)Q(a_2+1)}
+ \frac{P(a_3)P(a_3+1)}{4Q(a_3)Q(a_3+1)} } + \right.
$$ $$
+ \underline{\underline{\frac{P(a_1)P(a_1-1)}{4Q(a_1)Q(a_1-1)}
+ \frac{P(a_2)P(a_2-1)}{4Q(a_2)Q(a_2-1)}
+ \frac{P(a_3)P(a_3-1)}{4Q(a_3)Q(a_3-1)} }} +
$$ $$
\left. + \frac{P(a_1)P(a_2)}
{Q_{12}(a_1,a_2)}
+ \frac{P(a_2)P(a_3)}{Q_{23}(a_2,a_3)}
+ \frac{P(a_3)P(a_1)}{Q_{13}(a_1,a_3)}
\right)
$$
where $Q_{ij}(a_i,a_j) = Q_{ji}(a_j,a_i)
= (a_{ij}^2-\epsilon_+^2)(a_{ij}^2-\epsilon_-^2)
\prod_{k\neq i,j} a_{ik}(a_{ik}+\epsilon)a_{jk}(a_{jk}+\epsilon)$.

\bigskip

Matching with (\ref{CBw3c2}) at the first level defines
the AGT relation between the dimensions
\be
\Delta = \alpha^2+\beta^2, \ \ \ \ \ \
w = \alpha(\alpha^2-3\beta^2)
\label{alpar}
\ee
and the Nekrasov parameters $a_1,a_2,a_3=-a_1-a_2$. As usual,
in terms of the $\alpha$-parametrization, the AGT is a linear relation
(defined modulo rotations and Weyl reflections):
\be
{\rm AGT:}\ \ \ \ \ \ \ \ \ \ \ \ \
a_1 = \frac{\alpha}{\sqrt{3}}-\beta,\ \ \ \ \
a_2 = \frac{\alpha}{\sqrt{3}}+\beta,\ \ \ \ \
a_3 = -\frac{2\alpha}{\sqrt{3}}
\ee
Then,
\be
\sum_{i=1}^3 \frac{a_i^6}{\prod_{j\neq i}(a_i-a_j)^2}
= \frac{(\alpha^2+\beta^2)
(\alpha^6+75\alpha^4\beta^2-45\alpha^2\beta^4+9\beta^6)}
{2\beta^2(\beta^2-3\alpha^2)^2}
= \frac{\Delta\left(\Delta^3-\frac{8}{9}w^2\right)}{2(\Delta^3-w^2)}
\ee
i.e. the level one term in (\ref{NekSU3}) nicely reproduces
the one in (\ref{CBw3c2}) with $P(a)=a^6$, as claimed in \cite{Wyl,mmAGT}.

Now we can substitute the same values of $a_i$ and $P(a)$
into the second level term in (\ref{NekSU3}) and express the
result back through $\Delta$ and $w$ with the help of (\ref{alpar}).
Surprisingly or not, the latter step turns out possible.
Moreover, the answer \cite{mmAGT}%\cite[eq.(???)]{mmAGT}
is literally the same as in (\ref{CBw3c2}),
in perfect accordance with the AGT conjecture.
We do not need to include any $U(1)$ factors in this simplified
calculation, because we keep external momenta zero.
Switching on $\epsilon\neq 0$ ($c\neq 2$) and external momenta
is a more difficult calculation, requiring also some knowledge
from representation theory of $W^{(3)}$ algebras
(summarized for this purpose in \cite{MMMM}), but this
tedious job \cite{mmAGT} gives nothing new: the AGT relation
remains to be true.
Now there seem to be {\bf no room for doubt that the AGT conjecture is
valid}, time is now to learn lessons from it and apply it
to resolution of old and new problems.

\section{Some other comments on non-Virasoro AGT relations}

To conclude this note, we briefly comment on another kind
of problem \cite{Wyl} with the $W^{(3)}$ conformal blocks,
which appears for non-vanishing external momenta.

First, unlike the Virasoro case, the triple vertices for
$W^{(3)}$-descendants are not fully defined by the
$W$-symmetry alone: all the triple correlators of the form
$\langle V_1\ V_2\ (W_{-1}^kV_3)\rangle$  remain free
parameters. One can resolve this problem in two ways:
either via specifying a concrete conformal model, like
free fields or affine Toda, or
imposing restrictions on possible choice of the external states,
like the {\it speciality} conditions  of \cite{FLit},
requiring two of the four external states
($m-2$ in an $m$-point conformal block) to be the $W$-null-vectors
at level one.

Second, there is a mismatch between the number of free parameters
in conformal block and in the  $SU(3)$ Nekrasov function.
The structure of this mismatch is better seen if one considers
an arbitrary $N>2$.
In the $SU(N)$ case, there are $3N$ parameters in the Nekrasov
function ($N-1$ $a$'s plus $2N$ $\mu$'s plus $2$ $\epsilon$'s minus
$1$ common rescaling) and $5N-4$ parameters in the conformal block
($4$ external and $1$ internal $N-1$-component momenta plus
$1$ central charge). Thus the mismatch is: extra $2(N-2)$ parameters
on the CFT side.

If one tries to resolve the problem by considering the
free field model, then  one external and one internal momentum
are fully defined by the $3$ external momenta, so that the
number of parameters on the CFT side decreases to $3(N-1)+1=3N-2$
and this is a slight overplay: there are extra $2$ parameters on
the Nekrasov side.

An exact matching in the number of parameters is achieved
if one restricts to the {\it special} states \cite{Wyl}.
For $N=3$ this subtracts $2$ out of $5N-4=11$ parameters
on the CFT side what brings this number down to $9=3N$,
exactly the same as needed for the $SU(3)$ Nekrasov functions.

The problem, however, persists. Not only a selected set
of conformal blocks, but {\it all} of them can be one day
calculated in a given conformal model, as they can,
for example, in the model of $N-1$ free fields. The
$W^{(N)}$-symmetry does not fix them unambiguously,
but in a given model they are {\it all} well defined.
Nothing forbids these conformal blocks to have more free
parameters than there are available on the Nekrasov side of the
AGT relation. This does not happen to free fields, as we saw,
but a mismatch seems to exist already in the affine-Toda model.
Even for the free fields there is an open problem: the Nekrasov functions
describe this case under speciality conditions, but what happens if they are
lifted?
In any case, this mismatch seems certain to occur
in a generic model with the $W^{(N)}$-symmetry.

It is a very interesting and conceptually
important question, if the Nekrasov functions would still
provide a basis?
If not, should their set be somehow extended?
Do they really provide an exhaustive basis for expansion
of arbitrary generalized hypergeometric integrals from
\cite{SheVa}?
What the mismatch, if any should mean from the point
of view of the Dotsenko-Fateev approach?
Further work on the AGT relation will hopefully provide
answers to all these puzzles.

\section*{Acknowledgements}

We are indebted for hospitality and support to Prof.T.Tomaras and
the Institute of Theoretical and Computational Physics of
University of Crete, where part of this work was done.

The work was partly supported by Russian Federal Nuclear Energy
Agency and by RFBR grants 07-02-00878 (A.Mir.),
and 07-02-00645 (A.Mor.).
The work was also partly supported by grant FP7-REGPOT-1 (Crete HEP Cosmo 228644),
by joint grants 09-02-90493-Ukr,
09-02-93105-CNRSL, 09-01-92440-CE, 09-02-91005-ANF, INTERREG IIIA
(Greece-Cyprus) and by Russian President's Grant
of Support for the Scientific Schools NSh-3035.2008.2.

\end{document}